\documentclass[twocolumn,showpacs,amsmath,amssymb,aps,prl]{revtex4-1}
\usepackage{amsmath,amsfonts,bm}
\usepackage{graphicx}
\usepackage[colorlinks, linkcolor={red},citecolor={blue}]{hyperref}

\begin{document}
\title{Decoupling gravitational sources in general\\ relativity:
From perfect to anisotropic fluids}
\author{Jorge Ovalle}
\email[]{jovalle@usb.ve} \affiliation{Departamento de F\'{\i}sica, Universidad Sim\'on Bol\'{\i}var, 
AP 89000, Caracas 1080A, Venezuela\\The Institute for Fundamental Study, Naresuan University, Phitsanulok 65000, Thailand}
\begin{abstract}
We show the first simple, systematic and direct approach to decoupling gravitational sources in general relativity. As a direct application, a robust and simple way to generate anisotropic solutions for self-gravitating systems from perfect fluid solutions is presented.
\end{abstract}
\maketitle


\section{Introduction}
As is well known, in most cases solving Einstein field equations is a difficult task. Indeed, it is hard to obtain analytical solutions having some physical relevance, except for some specific situations \cite{Stephani}. One of these parti\-cular cases is the spherically symmetric space-time with a perfect fluid $\hat{T}_{\mu\nu}$ as a gravitational source~\cite{lake1,lake2,visser1}. However, as soon as the perfect fluid is coupled to complex forms of matter-energy to describe more realistic scenarios, namely, 
\begin{equation}
\label{a}
{T}_{\mu\nu}=\hat{T}_{\mu\nu}+\alpha\,\theta_{\mu\nu}\ ,
\end{equation}
with $\alpha$ a coupling constant and $\theta_{\mu\nu}$ any other form of gravitational source, then the situation changes radically, making it almost impossible to obtain analytical results that can be easily interpreted (for the case of anisotropic sources, see for instance~\cite{lake3,luis1}). In this respect, the so-called Minimal Geometric Deformation (MGD), originally proposed~\cite{jo1,jo2} in the context of the Randall-Sundrum brane-world~\cite{lisa1,lisa2} and extended to investigate new black hole
solutions~\cite{MGDextended1,MGDextended2}, has been successfully used to generate brane-world configurations from general relativistic perfect fluid solutions. Even exact and physically acceptable solutions for interior stellar distributions, which is a difficult task due to the existence of non-linear terms in the matter fields, were successfully generated~\cite{jo8}. The approach works very well, but the reason for this is, so far, unknown (for some recent applications, see for instance Refs.~\cite{jo11,jo12,roldaoGL,rrplb}). Nonetheless, we believe there must be a fundamental reason explaining all of this. One of the purposes of this paper is to explain this fundamental reason, as well as to show the potential of the MGD to be exploited in other areas beyond the brane-world. We will show that under the MGD lies a powerful and direct way of dealing with Einstein field equations, as described below.
\par
Let us start with a rhetorical and naive question. Would it not be ideal to solve Einstein field equations by solving the field equations for each gravitational source individually? That is, we could find the metric $g_{\mu\nu}$, and both energy-momentum tensors $\hat{T}_{\mu\nu}$ and $\theta_{\mu\nu}$, not by solving
\begin{equation}
\label{einst}
G_{\mu\nu}=-k^2\,\left(\hat{T}_{\mu\nu}+\alpha\,\theta_{\mu\nu}\right)\ ;\,\,\,\,\,\,\,k^2=8\pi\ ,
\end{equation}
but 
\begin{equation}
\label{einsth}
\hat{G}_{\mu\nu}=-k^2\,\hat{T}_{\mu\nu}\ ,\,\,\,\, {\rm to\,\,find}\,\,\,\{\hat{g}_{\mu\nu}, \hat{T}_{\mu\nu}\}
\end{equation}
and then
\begin{equation}
\label{einsts}
G^{*}_{\mu\nu}=-k^2\,\theta^{*}_{\mu\nu}\ ,\,\,\,\, {\rm to\,\,find}\,\,\,\{{g}^*_{\mu\nu}, {\theta}^*_{\mu\nu}\}
\end{equation}
and finally, we could obtain the metric $g_{\mu\nu}$ in Eq.~(\ref{einst}) by a {\it simple} combination of the two metrics found by Eqs.~(\ref{einsth}) and (\ref{einsts}), namely, $\hat{g}_{\mu\nu}$ and ${g}^*_{\mu\nu}$. Obviously, this would be great since it would introduce an unprecedented simplification. However, the consensus has told us for years not only the impossibility of carrying out this alternative but also the absurdity of it, given the highly non-linear and complex structure of Einstein field equations~\cite{superposition}. In this paper, contrary to the general belief that has lasted for years, we will show that the decoupling of the gravitational sources, as suggested in Eqs.~(\ref{einst})-(\ref{einsts}), can be done in a simple and direct way, at least for the spherically symmetric and static case, thus opening a range of new possibilities in the search for solutions to Einstein field equations.
\section{Einstein equations}
\label{s2}
\par
Let us start from Einstein field equations 
\begin{equation}
\label{corr2}
R_{\mu\nu}-\frac{1}{2}\,R\, g_{\mu\nu}
=
-k^2\,T^{\rm (tot)}_{\mu\nu}\ ,
\end{equation}
with
\begin{equation}
\label{emt}
T^{\rm (tot)}_{\mu\nu} = T^{\rm (m)}_{\mu\nu}+\alpha\,\theta_{\mu\nu}
\end{equation}
where
\begin{equation}
\label{perfect}
T^{\rm (m)}_{\mu \nu }=(\rho +p)\,u_{\mu }\,u_{\nu }-p\,g_{\mu \nu }
\end{equation}
is the four-dimensional energy-momentum tensor of ordinary matter, described by
a perfect fluid with 4-velocity field $u^\mu$, density $\rho$ and isotropic pressure $p$. On the other hand, the term $\theta_{\mu\nu}$ in Eq.~(\ref{corr2}) is any additional gravitational source coupled with the perfect fluid by the constant $\alpha$ \cite{Matt}. The source $\theta_{\mu\nu}$ may contain new fields, like scalar, vector and tensor fields. Since the Einstein tensor is divergence free, the energy-momentum tensor $T^{\rm (tot)}_{\mu\nu}$ satisfies the conservation equation
\begin{equation}
\nabla_\nu\,T^{{\rm (tot)}{\mu\nu}}=0
\ .
\label{dT0}
\end{equation}
\if
which yields to the conservation equation for both energy momentum-tensors, namely, the one for ordinary matter and the one associated with the new gravitational sector, hence
\begin{equation}
\nabla_\nu\,T^{{\rm (m)}{\mu\nu}}=0\ ;\,\,\,\,\,\, \nabla_\nu\,\theta^{\mu\nu}=0 \ .
\label{dT02}
\end{equation}
Other possibility satisfying the conservation equation in Eq. (\ref{dT0}) is the exchange of energy between the standard material fields and the new gravitational sector, that is
\begin{equation}
\label{exchange}
\nabla_\nu\,T^{{\rm (m)}{\mu\nu}} = -\alpha\nabla_\nu\,\theta^{\mu\nu}\ .
\end{equation}
\fi
\par 
In Schwarzschild-like coordinates, the spherically symmetric metric reads 
\begin{equation}
ds^{2}
=
e^{\nu (r)}\,dt^{2}-e^{\lambda (r)}\,dr^{2}
-r^{2}\left( d\theta^{2}+\sin ^{2}\theta \,d\phi ^{2}\right)
\ ,
\label{metric}
\end{equation}
where $\nu =\nu (r)$ and $\lambda =\lambda (r)$ are functions of the areal
radius $r$ only, ranging from $r=0$ (the star's center) to some $r=R$ (the
star's surface), and the fluid 4-velocity field is given by
$u^{\mu }=e^{-\nu /2}\,\delta _{0}^{\mu }$ for $0\le r\le R$.
The metric~(\ref{metric}) must satisfy the Einstein equations~(\ref{corr2}),
which explicitly read
\begin{eqnarray}
\label{ec1}
&&
-k^2
\left( \rho
+\alpha\,\theta_0^{\,0}
\right)
=-
\strut\displaystyle\frac 1{r^2}
+e^{-\lambda }\left( \frac1{r^2}-\frac{\lambda'}r\right)\ ,
\\
&&
\label{ec2}
-k^2
\strut\displaystyle
\left(-p+\alpha\,\theta_1^{\,1}\right)
=
-\frac 1{r^2}+e^{-\lambda }\left( \frac 1{r^2}+\frac{\nu'}r\right)\ ,
\\
&&
\label{ec3}
-k^2
\strut\displaystyle
\left(-p+\alpha\,\theta_2^{\,2}\right)
=
\frac 14e^{-\lambda }\left[ 2\,\nu''+\nu'^2-\lambda'\,\nu'
+2\,\frac{\nu'-\lambda'}r\right]
\ ,\nonumber
\\
\end{eqnarray}
while the conservation equation~(\ref{dT0}), which is a linear combination of Eqs.~(\ref{ec1})-(\ref{ec3}), yields
\begin{equation}
\label{con1}
-p'-\strut\displaystyle\frac{\nu'}{2}(\rho+p)+\alpha(\theta_1^{\,\,1})'-\strut\displaystyle\frac{\nu'}{2}\alpha(\theta_0^{\,\,0}-\theta_1^{\,\,1})-\frac{2\alpha}{r}(\theta_2^{\,\,2}-\theta_1^{\,\,1}) = 0
\ ,
\end{equation}
where $f'\equiv \partial_r f$.
We then note that the perfect fluid equations are formally
recovered for $\alpha\to 0$. By simple inspection of the field equations~(\ref{ec1})-(\ref{ec3}), we
can identify an effective density $\tilde{\rho}
=
\rho
+\alpha\,\theta_0^{\,0}$, an effective isotropic pressure $\tilde{p}_{r}
=
p-\alpha\,\theta_1^{\,1}$, and an effective tangential pressure $\tilde{p}_{t}
=
p-\alpha\,\theta_2^{\,2}
\ .$
This clearly illustrates that the source $\theta_{\mu\nu}$ generates an anisotropy $
\Pi
\equiv
\tilde{p}_{r}-\tilde{p}_{t}
=
\alpha\,(\theta_2^{\,2}-\theta_1^{\,1})$ inside the stellar distribution. At this stage the system~(\ref{ec1})-(\ref{ec3}) can be treated as an anisotropic fluid \cite{Luis}, and therefore, we would deal with five unknown functions, namely, the two metric functions $\nu(r)$ and $\lambda(r)$, and the effective functions $\tilde{\rho}$, $\tilde{p}_{r}$ and $\tilde{p}_{t}$.  However, we implement an unconventional way, as explained further below. 
\section{Minimal Geometric Deformation}
\label{s3}
Now let us implement the MGD to solve the system~(\ref{ec1})-(\ref{con1}). We will see that under this approach, the system will be transformed in such a way that the equations of motion associated with the source $\theta_{\mu\nu}$ will satisfy an effective ``quasi-Einstein system" [see further Eqs. (\ref{ec1d})-(\ref{ec3d})]. Let us start by considering a solution to the system~(\ref{ec1})-(\ref{con1}) with $\alpha=0$, namely, a GR perfect fluid solution $\{\xi,\mu,\rho,p\}$, where $\xi$ and $\mu$ are the new metric functions in  Eq.~(\ref{metric}), which now reads
\begin{equation}
ds^{2}
=
e^{\xi (r)}\,dt^{2}-\mu(r)^{-1}\,dr^{2}
-r^{2}\left( d\theta^{2}+\sin ^{2}\theta \,d\phi ^{2}\right)
\ ,
\label{pfmetric}
\end{equation}
where 
\begin{equation}
\label{standardGR}
\mu(r)
\equiv
1-\frac{k^2}{r}\int_0^r x^2\,\rho\, dx
=1-\frac{2\,m(r)}{r}
\end{equation}
is the standard GR solution containing the mass function $m$. Now let us turn on the parameter $\alpha$ to consider the effects of the source $\theta_{\mu\nu}$ on the perfect fluid solution $\{\xi,\mu\,\rho,p\}$. These effects can be encoded in the geometric deformation undergone by the perfect fluid geometry $\{\xi,\mu\}$ in Eq.~(\ref{pfmetric}), namely,
\begin{eqnarray}
\label{gd1}
\xi &\rightarrow &\nu\,=\,\xi+\alpha\,g\ ,
\\
\label{gd2}
\mu &\rightarrow &e^{-\lambda}=\mu+\alpha\,f ,
\end{eqnarray}
where $f$ and $g$ are, respectively, the geometric deformations undergone by the radial and temporal metric components. Of all the possibilities contained in Eqs.~(\ref{gd1}) and (\ref{gd2}), there is a specific one, the so-called minimal geometric deformation, for which 
\begin{eqnarray}
\label{gd11}
g&\rightarrow &\,0\
\\
\label{gd22}
f&\rightarrow &\,f^*\ .
\end{eqnarray}
The metric in Eq.~(\ref{pfmetric}) is thus minimally deformed by $\theta_{\mu\nu}$ and its radial metric component becomes
\begin{eqnarray}
\label{expectg}
\mu(r)\rightarrow\,e^{-\lambda(r)}
=
\mu(r)+\alpha\,f^{*}(r)
\ ,
\end{eqnarray}
while the temporal metric component $e^{\nu}$ remains unchanged [actually, $\nu(r)$ becomes $\nu(r,\alpha)$, after imposing matching conditions]. We want to emphasize that the expression in Eq.~(\ref{expectg}) is a linear decomposition of the inverse radial metric component $g^{11}$ in terms of a pure perfect fluid sector plus a contribution from the source $\theta_{\mu\nu}$. Now let us plug the  decomposition in Eq.~(\ref{expectg}) into the Einstein equations~(\ref{ec1})-(\ref{ec3}). The system is thus separated into two sets: (i) one having the standard Einstein field equations for a perfect fluid ($\alpha = 0$), whose metric is given by Eq.~(\ref{pfmetric}) with $\xi(r) = \nu(r)$; and (ii) one for the source $\theta_{\mu\nu}$, which reads
\begin{eqnarray}
\label{ec1d}
&&
-k^2\,\theta_0^{\,0}
=
\strut\displaystyle\frac{f^{*}}{r^2}
+\frac{f^{*'}}{r}\ ,
\\
&&
\label{ec2d}
-k^2
\strut\displaystyle
\,\theta_1^{\,1}
= f^{*}\left(\frac{1}{r^2}+\frac{\nu'}{r}\right)\ ,
\\
&&
\label{ec3d}
-k^2
\strut\displaystyle\,\theta_2^{\,2}
=\frac{f^{*}}{4}\left(2\nu''+\nu'^2+2\frac{\nu'}{r}\right)+\frac{f^{*'}}{4}\left(\nu'+\frac{2}{r}\right)
\ .
\end{eqnarray}
Since the Einstein tensor $\hat{G}_{\mu\nu}$ associated with the geometry of the perfect fluid, namely, $(\nu,\mu)$, must satisfy its respective Bianchi identity, the energy-momentum tensor $T^{\rm (m)}_{\mu \nu }$ in Eq. (\ref{perfect}) is conserved, and as a  consequence, the conservation equation in Eq.~(\ref{dT0}) yields $\nabla_\nu\,\theta^{\mu\nu}=0$, which explicitly reads
\begin{equation}
\label{con1d}
(\theta_1^{\,\,1})'-\strut\displaystyle\frac{\nu'}{2}(\theta_0^{\,\,0}-\theta_1^{\,\,1})-\frac{2}{r}(\theta_2^{\,\,2}-\theta_1^{\,\,1}) = 0
\ .
\end{equation}
Under these conditions, there is no exchange of energy-momentum between the perfect fluid and the source $\theta_{\mu\nu}$; their interaction is purely gravitational. 
\par 
There are some important features regarding the system~(\ref{ec1d})-(\ref{con1d}). First of all, it looks very similar to the standard spherically symmetric Einstein field equations for an anisotropic system with energy-momentum tensor $\theta_{\mu\nu}\,$; $\{{\rho}
= \theta_0^{\,0};\,\, {p}_{r}=-\theta_1^{\,1};\,\, {p}_{t}=-\theta_2^{\,2}\}$ and its respective conservation equation. However, it cannot be formally identified as the spherically symmetric Einstein field equations with radial metric component $f^{*}$ since the right-hand sides in Eqs. (\ref{ec1d}) and (\ref{ec2d}) do not have the standard expression for the Einstein tensor components 
$G_0^{\,\,0}$ and $G_1^{\,\,1}$ [there is a missed $-1/r^2$ in both right-hand sides in Eqs.~(\ref{ec1d}) and (\ref{ec2d})]. Despite the above, the system~(\ref{ec1d})-(\ref{con1d}) may be formally identified as Einstein equations for an anisotropic system with energy-momentum tensor ${\theta}^*_{\mu\nu}\,$ defined as
\begin{equation}
\label{shift2}
k^2\,\theta_\mu^{*\,\,\nu}=k^2\,\theta_\mu^{\,\nu}+\frac{1}{r^2}\left(\delta_\mu^{\,\,\,0}\,\delta_0^{\,\,\,\nu}+\delta_\mu^{\,\,\,1}\,\delta_1^{\,\,\,\nu}\right)\ ,
\end{equation}
with conservation equation 
\begin{equation}
\label{con1dd}
({\theta}_1^{*\,1})'-\strut\displaystyle\frac{\nu'}{2}({\theta}_0^{*\,0}-{\theta}_1^{*\,1})-\frac{2}{r}({\theta}_2^{*\,2}-{\theta}_1^{*\,1}) = 0
\ ,
\end{equation}
and metric 
\begin{equation}
ds^{2}
=
e^{\nu (r)}\,dt^{2}-\frac{dr^{2}}{f^{*}(r)}
-r^{2}\left( d\theta^{2}+\sin ^{2}\theta \,d\phi ^{2}\right)
\ .
\label{metricaniso}
\end{equation} 

\par
We want to mention an additional feature regarding the system~(\ref{ec1d})-(\ref{con1d}). As is well known, the Bianchi identity has trivial information as longer as no constraint has been imposed on the space-time geometry. Since the MGD imposes a kind of constraint through the expression in Eq. (\ref{expectg}), we should expect non-trivial information from the Bianchi identity. Indeed, the system~(\ref{ec1d})-(\ref{ec3d}) has two equations without the standard Einstein tensor components, and therefore we should anticipate that the conservation equation (\ref{con1d}) for the source $\theta_{\mu\nu}$ is no longer a linear combination of Eqs.~(\ref{ec1d})-(\ref{ec3d}). Remarkably, and despite the above, the conservation equation in Eq. (\ref{con1d}) still remains a linear combination of the system~(\ref{ec1d})-(\ref{ec3d}). As a consequence, under the MGD, we start with the indefinite system~(\ref{ec1})-(\ref{ec3}) and we end up with the set of equations for a perfect fluid $\{\nu,\mu,\rho,p\}$ plus a much simpler system of four unknown functions $\{f^{*},\,\theta_0^{\,0},\,\theta_1^{\,1},\,\theta_2^{\,2}\}$ satisfying three equations~(\ref{ec1d})-(\ref{ec3d}) (at this stage we suppose that we have already found a perfect fluid solution, thus $\nu$ is determined). In summary, the system~(\ref{ec1})-(\ref{ec3}) has been successfully decoupled into two systems, as suggested by Eqs.~(\ref{einst})-(\ref{einsts}). At this stage, a natural question arises: What happens if we consider an additional source $\Psi_{\mu\nu}$ in Eq.~(\ref{einst})? Namely, what if
\begin{equation}
\label{einst2}
G_{\mu\nu}=-k^2\,\left(\hat{T}_{\mu\nu}+\alpha\,\theta_{\mu\nu}+\beta\,\Psi_{\mu\nu}\right)\ ,
\end{equation}
with $\beta$ a coupling constant. The reader can easily follow the same scheme to find a successful decoupling by
\begin{eqnarray}
\label{expectg2}
e^{-\lambda(r)}
=
\mu(r)+\alpha\,f^{*}(r)+\beta\,h^*(r)
\ ,
\end{eqnarray}
where, in addition to Eqs.~(\ref{einsth}) and (\ref{einsts}), we have
\begin{equation}
\label{einsts2}
\tilde{G}_{\mu\nu}=-k^2\,\tilde{\Psi}_{\mu\nu}\ ,\,\,\,\, {\rm to\,\,find}\,\,\,\{\tilde{g}_{\mu\nu}, \tilde{\Psi}_{\mu\nu}\}\ ,
\end{equation}
with the metric $\tilde{g}_{\mu\nu}$ in Eq.~(\ref{einsts2}) given by
\begin{equation}
ds^{2}
=
e^{\nu (r)}\,dt^{2}-\frac{dr^{2}}{h^{*}(r)}
-r^{2}\left( d\theta^{2}+\sin ^{2}\theta \,d\phi ^{2}\right)\ ,
\label{metricaniso2}
\end{equation}
with the sources ${\Psi}_{\mu\nu}$ and $\tilde{\Psi}_{\mu\nu}$ related by the same expression as that in Eq.~(\ref{shift2}) for ${\theta}_{\mu\nu}$ and ${\theta}^*_{\mu\nu}$. We can see that this approach represents a {\it linear scheme} for decoupling gravitational sources. It can be summarized as follows: Given a static spherically symmetric perfect fluid whose energy-momentum tensor $\hat{T}_{\mu\nu}$ is coupled to $n$ gravitational sources $T_{\mu\nu}^{(i)}$, namely,
\begin{equation}
\label{MGD1}
T_{\mu\nu}=\sum_{i=0}^{n}\,\alpha_i\,T_{\mu\nu}^{(i)}\ ;\,\,\,\alpha_0=1\ ;\,\,\,T_{\mu\nu}^{0}=\hat{T}_{\mu\nu}\ ,
\end{equation}
the diagonal metric $g_{\mu\nu}$, the solution of the Einstein equation $G_{\mu\nu}=-k^2\,T_{\mu\nu}$, will be given by
\begin{eqnarray}
g_{\mu\nu}=\hat{g}_{\mu\nu}={g}_{\mu\nu}^{(i)}\ ;\,\,\mu=\nu\neq\,1\ , \\
g^{11}=\hat{g}^{11}+\alpha_1\,g^{11 (1)}+...+\alpha_n\,g^{11 (n)}\ .
\end{eqnarray}
This metric $g_{\mu\nu}$ is found by first solving Einstein field equations for the perfect fluid source $\hat{T}_{\mu\nu}$ 
\begin{equation}
\hat{G}_{\mu\nu}=-k^2\,\hat{T}_{\mu\nu}\ ;\,\,\,\,\,\,\,\, \nabla_\nu\,\hat{T}^{{\mu\nu}}=0\ ,
\end{equation}
and then by solving the remaining  $n$ ``quasi-Einstein equations" for the sources $T_{\mu\nu}^{(i)}$, namely
\begin{eqnarray}
\tilde{G}_{\mu\nu}^{(1)}&=&-k^2\,{T}_{\mu\nu}^{(1)}\ ;\,\,\,\,\, \nabla_\nu\,{T}^{{(1)\mu\nu}}=0\ ,
\nonumber \\
&\vdots &
\nonumber \\
\tilde{G}_{\mu\nu}^{(n)}&=&-k^2\,{T}_{\mu\nu}^{(n)}\ ;\,\,\,\,\, \nabla_\nu\,{T}^{{(n)\mu\nu}}=0\ ,
\label{MGD2}
\end{eqnarray}
where the divergence-free ``quasi-Einstein" tensor $\tilde{G}_{\mu\nu}$ and the standard one ${G}_{\mu\nu}$ are related by 
\begin{equation}
\tilde{G}_{\mu}^{\,\,\,\,\nu}={G}_{\mu}^{\,\,\,\,\nu}+\Gamma_{\mu}^{\,\,\,\,\nu}(g)\ ,
\end{equation}
with $\Gamma_{\mu}^{\,\,\,\,\nu}(g)$ a tensor that depends exclusively on ${g}_{\mu\nu}$. In our spherically symmetric representation it reads
\begin{equation}
\Gamma_{\mu}^{\,\,\,\,\nu}=\frac{1}{r^2}\left(\delta_\mu^{\,\,\,0}\,\delta_0^{\,\,\,\nu}+\delta_\mu^{\,\,\,1}\,\delta_1^{\,\,\,\nu}\right)\ .
\end{equation}
The explicit components of $\tilde{G}_{\mu}^{\,\,\,\,\nu}$ in terms of the metric in Eq.~(\ref{metricaniso}) are shown in the right-hand side of Eqs.~(\ref{ec1d})-(\ref{ec3d}). 
\section{Interior: from perfect to anisotropic fluids}
\label{s5}
\par
In order to see the robustness of the MGD, let us solve the Einstein field equations in Eqs.~(\ref{ec1})-(\ref{ec3}) for the interior of a self-gravitating system. The first step is to turn off $\alpha$ in order to find a solution for the perfect fluid. Instead of really solving this sector, we simply choose an already-known solution with physical relevance, for instance, the well-known Tolman IV solution $(\nu,\mu,\rho, p)$ for perfect fluids, namely, 
\begin{equation}\label{tolman00}
e^{\nu}=B^2\,\left(1+\frac{r^2}{A^2}\right)\ ,
\end{equation}
\begin{equation}\label{tolman11}
\mu=\frac{\left(1-\frac{r^2}{C^2}\right)\left(1+\frac{r^2}{A^2}\right)}{1+\frac{2\,r^2}{A^2}}\ ,
\end{equation}
\begin{equation}\label{tolmandensity}
\rho(r) =\frac{3A^4+A^2\left(3C^2+7r^2\right)+2 r^2 \left(C^2+3 r^2\right)}{k^2\,C^2\left(A^2+2r^2\right)^2}\ ,
\end{equation}
and
\begin{equation}
\label{tolmanpressure} p(r)=\frac{C^2-A^2-3r^2}{k^2\,C^2\left(A^2+2r^2\right)}\ .
\end{equation}
The constants $A$, $B$ and $C$ in Eqs.~(\ref{tolman00})-(\ref{tolmanpressure}) are found by matching conditions, yielding $A^2/R^2=\frac{1-3\,c}{c}$; $B^2=1-3\,c$ and $C^2/R^2=c^{-1}$, with $c\equiv\,M_0/R<4/9$ and $M_0$ the total mass $m(R)$ in Eq.~(\ref{standardGR}). Now let us turn on $\alpha$ to find the metric in Eq.~(\ref{metric}), which is the solution of Eqs.~(\ref{ec1})-(\ref{ec3}). The temporal and radial metric components are given by Eqs.~(\ref{tolman00}) and~(\ref{expectg}), respectively, while the deformation $f^*(r)$ and the source $\theta_{\mu\nu}$ are found through Eqs.~(\ref{ec1d})-(\ref{ec3d}). Hence, we need to provide additional information to close the system~(\ref{ec1d})-(\ref{ec3d}). We have many alternatives, either an equation of state associated with the source $\theta_{\mu\nu}$ or some physically motivated restriction on $f^*(r)$. In any case, we must be careful in keeping the physical acceptability of our solution, which is not a trivial matter. Regarding this, we can take advantage of the structure of the system~(\ref{ec1d})-(\ref{ec3d}), namely, the (quasi) Einstein equations for the source ($\theta_{\mu\nu}$) $\theta^*_{\mu\nu}$. From Eq.~(\ref{ec2d}) we can see that the following choice for $f^*(r)$,
\begin{equation}
\label{mimic}
f^*(r)=-\mu(r)+\frac{1}{1+r\,\nu'(r)}
\end{equation}
yields
\begin{equation}
\label{mimic2}
k^2
\strut\displaystyle
\,\theta_1^{\,1}
= -\frac{1}{r^2}+\mu(r)\left(\frac{1}{r^2}+\frac{\nu'}{r}\right)\ .
\end{equation}
As a consequence, the radial pressure $\theta_1^{\,1}$ in Eq.~(\ref{ec2d}) mimics the (physically acceptable) perfect fluid pressure $p(r)$ in Eq.~(\ref{ec2}). Therefore, after imposing the matching conditions $C^2=A^2+3R^2$, the effective radial pressure in Eq.~(\ref{ec2}) reads 
\begin{equation}
\label{pressmimic}
\tilde{p}_r=\frac{3\,(1-\alpha)(R^2-r^2)}{k^2\,(A^2+3\,R^2)(A^2+2\,r^2)}\ ,
\end{equation}
thus leading to a physically acceptable anisotropic fluid solution $\{\tilde{\rho},\,\tilde{p}_r,\,\tilde{p}_t\}$ to Eqs.~(\ref{ec1})-(\ref{ec3}). The ``mimic constraint" in Eq.~(\ref{mimic}) is the simplest way to extend the physical acceptability of the perfect fluid solution in the anisotropic domain. In this case $f^*(r)<0$; hence, it strengthens the gravitational field. Many interesting properties of this anisotropic exact solution can be studied. However, a complete and detailed analysis is beyond the objective of this paper. The reader can prove, by following this simple scheme, that any known perfect fluid solution can be consistently extended to generate new anisotropic solutions. In principle, for each perfect fluid solution there will be as many anisotropic solutions as independent constraints can be imposed on the system~(\ref{ec1d})-(\ref{ec3d}). We conclude by emphasizing that the approach described here represents an effective and systematic method for decoupling gravitational sources, as shown through Eqs.~(\ref{MGD1})-(\ref{MGD2}). In addition to other virtues, it is a robust and direct way to generate anisotropic solutions for self-gravitating systems from perfect fluid solutions. It represents, as far as we know, the first simple, systematic and direct method that shows how to decouple gravitational sources in general relativity, and therefore a good guide to consider more complex scenarios. In this respect, the method can be generalized by also considering a deformation on the temporal metric component, and thus to investigate the impact of different gravitational sources, generically represented here as $\theta_{\mu\nu}$: for instance, to consider the Maxwell tensor or the coupling with Klein-Gordon scalar fields, or even both simultaneously. On the other hand, this approach will simplify the analysis of the stability of self-gravitating systems, which can be developed sector by sector under the MGD. In this respect, for instance, we know that finding analytic and physically relevant solutions for the interior of a self-gravitating system minimally coupled to a scalar field $\psi$ seems an impossible task to carry out. However, under the MGD, we can start with an exact and physically acceptable perfect fluid solution, and then focus solely on the scalar sector represented by $\psi$. This obviously represents a great simplification. In addition to all of the above, since the interaction among the sources under the MGD is purely gravitational, it is particularly useful to study the interaction between ordinary matter and the conjectured dark matter. Finally, some questions regarding the MGD and the way it works remain open: for instance, its validity for time-dependent solutions and a possible extension beyond the spherical symmetry, as well as a formal mathematical description, if any, of the decoupling in Eqs.~(\ref{MGD1})-(\ref{MGD2}) in terms of Killing vectors.
\par
{
{\it 
\hspace{2cm}`` Yo amo los mundos sutiles,\\
\hspace*{2.6cm} ingr\'avidos y gentiles,\\
\hspace*{2.6cm} como pompas de jab\'on "
\vspace{0.1 cm}\\ \hspace*{2.8cm} Antonio Machado}}

%
%
%
%



\begin{thebibliography}{99}

\bibitem{Stephani} Hans Stephani, Dietrich Kramer, Malcolm Maccallum, Cornelius Hoenselaers, Eduard Herlt, Exact Solutions of Einstein’s Field Equations(Cambridge University Press, Cambridge,
2003).
%
\bibitem{lake1} M. S. R. Delgaty and K. Lake, 
Comput. Phys. Commun. {\bf 115} 395 (1998). 
%
\bibitem{lake2}  K. Lake, 
Phys.Rev. D {\bf 67} 104015 (2003). 
%
\bibitem{visser1} Petarpa Boonserm, Matt Visser, Silke Weinfurtner,  
Phys.Rev. D {\bf 71} (2005) 124037. 
%
\bibitem{lake3} K. Lake, 
Phys.Rev.Lett. {\bf 92} (2004) 051101.
%
\bibitem{luis1} L. Herrera, J. Ospino and A. Di Prisco, 
Phys. Rev. D {\bf 77}, 027502 (2008).
%
%
\bibitem{jo1} 
J. Ovalle, 
{\sl Mod. Phys. Lett. A}, {\bf 23}, 3247 (2008).
%
\bibitem{jo2} 
J. Ovalle, {\it Braneworld stars: anisotropy minimally projected onto the brane}, 
{\sl in Gravitation and Astrophysics}
(ICGA9), edited by J. Luo, (World Scientific, Singapore, 2010), pp. 173-
182.

\bibitem{lisa1} L. Randall and R. Sundrum, 
{\sl Phys. Rev. Lett}. {\bf 83}, 3370 (1999)

\bibitem{lisa2} L. Randall and R. Sundrum, 
{\sl Phys. Rev. Lett} {\bf 83}, 4690 (1999).

\bibitem{MGDextended1} Roberto Casadio, Jorge Ovalle, Roldao da Rocha, 
Class. Quantum Grav. {\bf 32}, 215020 (2015).
\bibitem{MGDextended2} J Ovalle, 
Int. J. Mod. Phys. Conf. Ser. {\bf 41} 1660132 (2016).
\bibitem{jo8} 
J. Ovalle, F. Linares, 
{\sl Phys. Rev. D}, {\bf 88}, 104026 (2013).
\bibitem{jo11} 
J. Ovalle, L.A. Gergely, R. Casadio, 
{\sl Class. Quantum Grav.}, {\bf 32}, 045015 (2015).
%
\bibitem{jo12}  R. Casadio, J. Ovalle, R. da Rocha, 
{\sl Europhys. Lett.}, {\bf 110}, 40003 (2015).
\bibitem{roldaoGL} R. T. Cavalcanti, A. Goncalves da Silva, Roldao da Rocha, 
Class. Quantum Grav. {\bf 33}, 215007 (2016).
%
\bibitem{rrplb} Roberto Casadio, Roldao da Rocha, 
Phys. Lett. B {\bf 763}, 434 (2016).
%
\bibitem{superposition} Some conditions are established by Basilis C. Xanthopoulos, {\it Superposition of solutions in general relativity} in Lecture Notes in Physics {\bf 239},
edited by R. Martini. Springer-Verlag, Berlin, 1985., p.109, under which Einstein equations behave like linear equations.
%
%
\bibitem{Matt} Piyabut Burikham, Tiberiu Harko, Matthew J. Lake, 
Phys. Rev. D {\bf 94}, 064070 (2016).
%
\bibitem{Luis} L. Herrera, N.O. Santos,  	
Phys.Rept. {\bf 286} 53 (1997).
%
\end{thebibliography}
\end{document}